\documentclass[fleqn,12pt]{wlscirep}

\usepackage{hyperref}         
\usepackage{url}              
\usepackage{bm}
\usepackage{siunitx}
\usepackage{soul,color}
\usepackage[export]{adjustbox}
\usepackage{lineno}

\title{Stress-dependent activation entropy in thermally activated cross-slip of dislocations}

\author[1]{Yifan Wang}
\author[1,*]{Wei Cai}
\affil[1]{Department of Mechanical Engineering, Stanford University, Stanford, CA, 94305, U.S.A.}

\affil[*]{Corresponding author's E-mail: caiwei@stanford.edu}

\begin{document}

\flushbottom
\maketitle
\thispagestyle{empty}


\noindent {\bf
Cross slip of screw dislocations in crystalline solids is a stress-driven thermally activated process essential to many phenomena during plastic deformation, including dislocation pattern formation, strain hardening, and dynamic recovery.
Molecular dynamics (MD) simulation has played an important role in determining the microscopic mechanisms of cross slip.
However, due to its limited timescale, MD can only predict cross-slip rates in high-stress or high-temperature conditions.
The transition state theory can predict the cross-slip rate over a broad range of stress and temperature conditions, but its predictions have been found to be several orders of magnitude too low in comparison to MD results.
This discrepancy can be expressed as an anomalously large activation entropy whose physical origin remains unclear.
Here we resolve this discrepancy by showing that the large activation entropy results from anharmonic effects, including thermal softening, thermal expansion, and soft vibrational modes of the dislocation.
We expect these anharmonic effects to be significant in a wide range of stress-driven thermally activated processes in solids.
}

%
\section*{Introduction}

Dislocation slip is the primary source of plastic deformation in crystalline solids.
Cross-slip occurs when a screw dislocation changes its slip plane (Fig.~1(a)). 
This stress-driven, thermally-activated process is critical in creating dislocation patterns \cite{jackson_dislocation_1985} and bypassing obstacles \cite{humphreys_deformation_1970, singh_atomistic_2011},
which leads to strain hardening and dynamic recovery \cite{johnston_dislocation_1960, ikeno_behavior_1972, caillard_chapter_2003} during plastic deformation.
It has long been challenging to accurately predict the cross-slip rate as a function of stress and temperature.  
Many experimental \cite{bonneville_cross-slipping_1979, bonneville_study_1988}
and theoretical \cite{puschl_calculation_1993, vegge_determination_2000}
analyses have been performed to determine the activation parameters for cross slip based on the continuum theory of dislocations.
However, the applicability of the continuum theory is questionable \cite{kang_stress_2014} since the changes in dislocation core structure during cross slip can be confined to only a few lattice spacings.
Fully atomistic models are needed to uncover the fundamental physical mechanisms of cross-slip.
Unfortunately, direct molecular dynamics (MD) simulation has a limited timescale (typically less than 100~ns), so it is only applicable when cross slip occurs at a high rate, i.e. under a high-stress or high-temperature condition~\cite{oren_kinetics_2017,esteban-manzanares_influence_2020}.

%
The transition state theory (TST), combined with minimum energy paths (MEP) calculations, provides a theoretical framework to predict the rate of thermally activated processes in solids over a wide range of stress and temperature conditions \cite{vineyard_frequency_1957, voter_method_1997, delph_harmonic_2013}. 
For a screw dislocation segment of length $L$, the cross-slip rate as a function of temperature $T$ under applied stress tensor $\bm{\tau}_{\rm app}$ (Fig.~1(b)) can be written as,

\begin{align}
    r(T,\bm{\tau}_{\rm app},L)=\nu(L)\,\exp\left[-\frac{H_{\rm c}(\bm{\tau}_{\rm app})}{k_{\rm B}T}\right]
    \label{eq:Arrhenius_rate}
\end{align}

\noindent where $H_{\rm c}$ is the activation enthalpy obtained from MEP calculations, and $k_{\rm B}$ is the Boltzmann constant.
The rate prefactor $\nu(L)$ is proportional to the dislocation length $L$ and can be written as $\nu(L)=\nu_{\rm e}\,L/b$ where $\nu_{\rm e}$ is an effective attempt frequency, and $b$ is the magnitude of the dislocation Burgers vector and hence the smallest repeat distance along the dislocation.
%
%
%
In cross-slip models used in discrete dislocation dynamics (DDD) simulations, the rate prefactor is linked to the vibrational frequency of the dislocation line, and is commonly expressed as $\nu(L)=\nu_{\rm D}\,L/L_0$
%
where $\nu_{\rm D}\sim\SI{e13}{s^{-1}}$ is the Debye frequency, and $L_0=\SI{1}{\micro\meter}$ is a reference length~\cite{kubin_dislocation_1992,hussein_microstructurally_2015}.
%
Given that the reported activation enthalpy $H_{\rm c}$ for the cross slip in Cu is in the range of
\SIrange[range-units=single, range-phrase={ -- }]{0.5}{3}{\electronvolt}
\cite{bonneville_cross-slipping_1979, bonneville_study_1988,puschl_calculation_1993, rasmussen_atomistic_1997},
together with the rate prefactor estimates above, cross slip is not expected to occur in direct MD simulations except at very high temperatures or stresses.
%

%
However, previous studies
\cite{vegge_determination_2000,oren_kinetics_2017,esteban-manzanares_influence_2020}
have shown that cross slip occurs in direct MD simulations at a much higher rate than expected (see Fig.~1(c)).
%
This discrepancy has led to the suggestion that the previous estimates of the rate prefactor is incorrect, and need to be multiplied by a factor of $\exp\left[\Delta S_{\rm c}(\bm{\tau}_{\rm app})/k_{\rm B} \right]$, where $\Delta S_{\rm c}$ is a stress-dependent activation entropy whose physical origin has remained elusive~\cite{ryu_entropic_2011,nguyen_atomistic_2011,saroukhani_harnessing_2016,proville_modeling_2020}.
It has been estimated either through an empirical estimate based on Meyer-Neldel rule~\cite{esteban-manzanares_influence_2020} or simplified line tension models~\cite{sobie_modal_2017}, but not from fully atomistic models due to numerical difficulties~\cite{vegge_determination_2000, proville_modeling_2020}.
%
%
%
%
%
%
%
%
%
The unknown origin of the activation entropy has raised doubts about whether TST is even applicable in thermally activated processes such as cross slip~\cite{saroukhani_harnessing_2016, saroukhani_investigating_2017}.

This work provides a systematic and fully atomistic approach to resolve the discrepancy in the cross slip rates and uncover the physical origin of the anomalously large activation entropy.
%
We carry out high-throughput minimum-energy paths (MEP) calculations to map out the stress dependence of the activation enthalpy $H_{\rm c}(\bm{\tau}_{\rm app})$.
The rate prefactor is determined from the harmonic transition state theory (HTST), 
with essential corrections applied to soft vibrational modes of the dislocation. 
Our approach reveals that in order to resolve the rate discrepancy between MD and TST predictions, anharmonic effects of thermal softening and thermal expansion must be appropriately considered.
%
%
%
%
These effects cause the solid to experience more significant shear and volumetric deformations when temperature increases at constant applied stress, and cause a pronounced drop in the cross-slip activation barrier, giving rise to the activation entropy $\Delta S_{\rm c}$. 
%
%
%
We find that $\Delta S_{\rm c}$ is more pronounced at higher stress, contrary to previous estimates~\cite{esteban-manzanares_influence_2020} based on the Meyer-Neldel rule~\cite{meyer_relation_1937}.
%
%
This work demonstrates the applicability of HTST (after corrections) to dislocation cross-slip and provides a quantitative approach to predict its rate and activation entropy. 
The significant activation entropy is expected to influence the rate of a wide range of stress-driven thermally-activated processes in solids, such as phase transformation and twin boundary migration.
%

\section*{Results}

We use face-centered cubic nickel as an example to investigate dislocation cross-slip behaviors.
%
The interatomic force field is modeled by the embedded-atom model (EAM) `vnih' \cite{rao_atomistic_1999}
because its stacking fault energy is in good agreement with both experimental measurements and first-principle calculations \cite{rao_calculations_2011,kang_stress_2014}.
%
The simulation cell is large enough ($N=\num{78400}$ atoms) to avoid boundary effects on dislocation cross slip rates.
%
A screw dislocation along the $x$-direction passes through the
center of the simulation cell.
The cell is periodic in $x$- and $z$- directions and has free surfaces on the $y$-direction.
Shear stresses $\bm{\tau}_{\rm app} = (\sigma_{\rm e}^{\rm g}, \sigma_{\rm s}^{\rm c}, \sigma_{\rm e}^{\rm c})$ are applied to provide driving force for cross-slip. 
As shown in Fig.~1(a),
the applied stress contains Escaig ($_{\rm e}$) and Schmid ($_{\rm s}$) components on the original slip ($^{\rm g}$) plane $(111)$ and the cross-slip ($^{\rm c}$) plane $(11\bar{1})$ (see Methods).
The Schmid stress on the original slip plane $\sigma_{\rm s}^{\rm g}$ is set to zero so that the dislocation does not move prior to cross-slip~\cite{kang_stress_2014, kuykendall_stress_2020, esteban-manzanares_influence_2020}.
%

MD simulations of cross-slip are carried out using the LAMMPS package \cite{thompson_lammps_2022}.
The initial configuration is heated up to the target temperature $T$ using the Nos\'{e}-Hoover thermostat (NVT ensemble) while keeping a constant applied stress at $\bm{\tau}_{\rm app} = (\num{-0.6}, \num{-0.8}, \num{0.8})\,\si{GPa}$ by adjusting the strain.
After equilibration, the simulation continues at constant $T$ and the corresponding stress until the dislocation cross-slips (at time $t_{\rm cs}$) and annhilates at the surface (see Methods).
%
%
The MD simulation is repeated 32 times at each temperature.
The cross-slip rate $r_{\rm MD}$, estimated as the inverse of the average cross-slip time $\bar{t}_{\rm cs}$, is plotted against the temperature in Fig.~1(c).
The temperature dependence of the cross-slip rate is seen to follow the Arrhenius law,

\begin{align}
    r_{\rm MD} 
        = \nu_{\rm MD}\,\exp\left[-\frac{H_{\rm c}^{\rm MD}}{k_{\rm B}T}
            \right]
\end{align}
%
where $H_{\rm c}^{\rm MD} = \SI{0.60}{eV}$ 
and $\nu_{\rm MD} = \SI{2.57e16}{s^{-1}}$ 
are parameters obtained from fitting the MD data.


We proceed to analyze the cross-slip rates by TST. The activation enthalpy $H_{\rm c}$ represents the energy difference between the transition state (i.e., saddle point on the potential energy landscape) and the initial state of the thermally activated cross-slip.
To find the transition state under the applied stress $\bm{\tau}_{\rm app}$, we first determine the minimum-energy path (MEP) using the free-end string method~\cite{e_string_2002,kuykendall_investigating_2015}.
%
Given the MEP, the exact transition state (saddle point) is then obtained by the dimer method~\cite{henkelman_dimer_1999} (see Methods).
Fig.~1(b) illustrates two converged MEPs with and without the applied stress $\bm{\tau}_{\rm app}$ corresponding to the MD simulations, respectively.
As expected, the applied stress lowers the activation enthalpy of cross slip.
Furthermore, the activation enthalpy of cross slip under the applied stress, $H_{\rm c}=\SI{0.60}{eV}$, perfectly matches the value $H_{\rm c}^{\rm MD}$ extracted from the MD simulations (Fig.~1(c)).
On the other hand, if we adopt the commonly used estimate for the frequency prefactor~\cite{kubin_dislocation_1992,hussein_microstructurally_2015}, $\nu(L)=\nu_{\rm D}\,L/L_0$, for the dislocation length ($L\approx\SI{10}{nm}$) considered here, we would arrive at $\nu(L) \approx 10^{11} \, {\rm s}^{-1}$, which is more than five orders of magnitude lower than MD predictions (see Fig.~1(c)).
This paper's primary purpose is to identify the physical origin of this discrepancy.
%
%
%
%



To go beyond a heuristic estimate, we use the harmonic transition state theory (HTST) to compute the rate prefactor more rigorously.
%
%
In HTST, the rate prefactor is expressed as follows~\cite{vineyard_frequency_1957},

\begin{align}
    \nu_{\rm HTST} = \frac{\prod_{i=1}^{3N-3}\,\nu_i^A}{\prod_{j=1}^{3N-4}\,\nu_j^S}
    \label{eq:nu_HTST}
\end{align}
where $\nu_i^A$ and $\nu_i^S$ are frequencies of the eigenmodes of the initial state (A) and the transition state (S), respectively.
The three rigid-body translational models (with zero frequency) are excluded from the product in both states A and S.
For state S, the mode along the reaction coordinate (with imaginary frequency) is also excluded.
%
%
Although HTST is often employed to study thermally activated processes in solids at moderately low temperatures,
it has never been successfully applied to dislocation cross-slip due to several challenges.

First, a direct implementation of Eq.~(\ref{eq:nu_HTST}) requires diagonalizing the Hessian matrix of the system to obtain the eigen-frequencies~\cite{proville_quantum_2012} (for both states A and S).
The Hessian matrix 
is quite large (size $3N\times3N$) and a full diagonalization is computationally very expensive.
In this work, we take advantage of the fact that the product of eigen-frequencies can be obtained from the determinant of the Hessian matrix, which can be computed much more efficiently (e.g. using LU decomposition) than to obtain all the eigen-frequencies individually. To avoid the determinant becoming zero due to the rigid-body translation modes, we slightly perturb the Hessian matrix to impart a small but non-zero frequency to these modes (see Methods).

Second, the harmonic approximation is not valid at room temperature or above for some of the \emph{soft vibrational modes}.
%
For example, the saddle state S contains a constriction of the stacking fault, which can be formed anywhere along the dislocation line.
Motion of this constriction along the dislocation line, i.e. the so-called Goldstone mode, produces periodic energy variations with an amplitude of around \SI{20}{meV}~\cite{vegge_determination_2000}, even lower than the thermal energy.
%
%
In this case, approximating the periodic potential landscape by a quadratic function leads to a large error in the partition function.
%
Here we account for these soft vibrational modes by numerically evaluating the partition function in their eigen-directions,
and introduce a correction factor (${\tilde{\nu}_{\rm A}}/{\tilde{\nu}_{\rm S}}$) to the cross-slip rate prediction,
where ${\tilde{\nu}_{\rm S}}$ is the correction factor for the Goldstone mode in the saddle state S, and ${\tilde{\nu}_{\rm A}}$ is the correction factor for the uniform glide mode of the screw dislocation on its slip plane in state A (see \hl{Supplementary Text I}).

Using the above two methods, we can now evaluate the HTST-based rate prefactor, $\nu(L) = \nu_{\rm HTST}\cdot \tilde{\nu}_{\rm A}/{\tilde{\nu}_{\rm S}}$.
For the stress condition considered above, $\nu(L) = \SI{7.73e12}{s^{-1}}$, which, although higher than previous estimates, is still much lower than $\nu_{\rm MD}$.
As a result, the predicted cross-slip rate (black line) is still $3$-$4$ orders of magnitude lower than the MD results (see Fig.~2(b)).

To resolve the remaining discrepancy, we note that the activation enthalpy $H_{\rm c}$ at a given stress $\bm{\tau}_{\rm app}$ is often computed as an activation energy $E_{\rm c}$ at a given strain $\bm{\varepsilon}$ corresponding to stress $\bm{\tau}_{\rm app}$.
To make this point more explicit, we express the cross-slip rate as a function of strain $\bm{\varepsilon}$ and temperature $T$,

\begin{align}
    r_{\rm HTST}(\bm{\varepsilon},T)= 
        \nu_{\rm HTST}
        \,\frac{\tilde{\nu}_{\rm A}}{\tilde{\nu}_{\rm S}}
        \exp\left[-\frac{E_{\rm c}(\bm{\varepsilon})}{k_{\rm B}T}\right]
    \label{eq:rHTST}
\end{align}
For consistency, $\bm{\varepsilon}$ should be the strain $\bm{\varepsilon}_T\equiv\bm{\varepsilon}(\bm{\tau}_{\rm app}, T)$ corresponding to stress $\bm{\tau}_{\rm app}$ at temperature $T$.
However, most of the MEP methods, which are based on energy minimization, are performed at zero temperature.
Let us define $\bm{\varepsilon}_0\equiv\bm{\varepsilon}(\bm{\tau}_{\rm app}, 0)$ as the strain corresponding to stress $\bm{\tau}_{\rm app}$ at zero temperature. 
In the above, we have reported that $E_{\rm c}(\bm{\varepsilon}_0) = H_{\rm c}(\bm{\tau}_{\rm app}) = \SI{0.60}{eV}$.
From Eq.~(\ref{eq:rHTST_DS}), it can be clearly seen that an inconsistency would arise if $\bm{\varepsilon} = \bm{\varepsilon}_T$ is used on the left hand side and $\bm{\varepsilon} = \bm{\varepsilon}_0$ is used on the right hand side.

While the difference between $\bm{\varepsilon}_T$ and $\bm{\varepsilon}_0$ has been implicitly assumed to be small and often neglected, here we show that it has a pronounced effect on the predicted cross-slip rate.
If the applied stress $\bm{\tau}_{\rm app}$ remains constant as temperature is increased, the strain $\bm{\varepsilon}_T$ increases in both the deviatoric and volumetric components, as sketched in the inset of Fig.~2(a).
Fig.~2(a) shows that the computed activation energy $E_{\rm c}(\bm{\varepsilon}_T)$ decreases linearly with temperature, i.e.,
$E_{\rm c}(\bm{\varepsilon}_T) = E_{\rm c}(\bm{\varepsilon}_0) - T\cdot \Delta S_{\rm c}(\bm{\tau}_{\rm app})$, where $\Delta S_{\rm c}(\bm{\tau}_{\rm app}) = 8.0\,k_{\rm B}$ is the negative slope of the $E_{\rm c}$-$T$ curve, and can be called an \emph{activation entropy}.
%
Inserting this expression of $E_{\rm c}(\bm{\varepsilon}_T)$ into Eq.~(\ref{eq:rHTST}), we can express the HTST-based rate prediction as,

\begin{align}
    r_{\rm HTST}(\bm{\varepsilon}_T,T)= 
        \nu_{\rm HTST}
        \,\frac{\tilde{\nu}_{\rm A}}{\tilde{\nu}_{\rm S}}
        \exp\left[\frac{\Delta S_{\rm c}(\bm{\tau}_{\rm app})}{k_{\rm B}}\right]
        \exp\left[-\frac{E_{\rm c}(\bm{\varepsilon}_0)}{k_{\rm B}T}\right]
    \label{eq:rHTST_DS}
\end{align}
The new rate prefactor, $\nu(L) = \nu_{\rm HTST}\cdot ({\tilde{\nu}_{\rm A}}/{\tilde{\nu}_{\rm S}}) \cdot \exp(\Delta S_{\rm c} / k_{\rm B}) = \SI{2.30e16}{s^{-1}}$, is in very good agreement with $\nu_{\rm MD}$.
Fig.~2(b) shows that the resulting HTST-based predictions of cross-slip rates now agree well with MD results.

\section*{Discussion}

In the example considered above, we observe that the large discrepancy between previous TST-based predictions of cross-slip rate and MD results is mostly due to the change of strain with increasing temperature at a constant applied stress.
Due to the thermal softening effect, the same shear stress will result in greater shear strain at higher temperature.
Due to the thermal expansion effect, the volumetric strain also increases with increasing temperature.
We have repeated the MD simulations and HTST calculations of cross-slip rates at two more applied stress conditions, and the results support the same conclusions (\hl{Supplementary Text II}).

To examine how does the activation entropy depends on the applied stress, we compute $\Delta S_{\rm c}$ at 27 different stress conditions (for $\sigma_{\rm e}^{\rm g}=0,-0.4,-0.8\,\si{GPa}$,
 $\sigma_{\rm s}^{\rm c}=0,-0.4,-0.8\,\si{GPa}$, and
 $\sigma_{\rm e}^{\rm c}=0, 0.4, 0.8\,\si{GPa}$, respectively).
We have previously shown that the activation enthalpy $H_c(\bm{\tau}_{\rm app})$ as a function of these three shear stress components can be expressed in terms of a one-dimension function of an \emph{effective stress}~\cite{kuykendall_stress_2020}, defined as $\tau^{*} = C_{\rm e}^{\rm g}\sigma_{\rm e}^{\rm g} + C_{\rm e}^{\rm c}\sigma_{\rm e}^{\rm c} + (D_{\rm s}^{\rm c}\sigma_{\rm s}^{\rm c})^2$, where $C_{\rm e}^{\rm g}$, $C_{\rm e}^{\rm c}$ and $D_{\rm s}^{\rm c}$ are fitting constants.
Fig.~3 shows that the activation entropy $\Delta S_{\rm c}$ generally increases with the effective stress $\tau^*$, although it is not a function of $\tau^*$ alone (see \hl{Supplementary Text II}).
The empirical Mayer-Neldel rule, $S_{\rm c}=H_{\rm c}/T_{\rm m}$, where $T_{\rm m}$ is the melting temperature, is often used to estimate the activation entropy~\cite{esteban-manzanares_influence_2020}.
Because the cross-slip activation enthalpy $H_{\rm c}(\bm{\tau}_{\rm app})$ is a monotonically decreasing function of $\tau^*$, it is clear that the Mayer-Neldel rule does not apply to cross-slip.
As shown in Fig.~3, $\Delta S_{\rm c}$ for cross slip becomes smaller at lower stress; in fact $\Delta S_{\rm c}$ vanishes in the zero stress limit, as we will show below.
This may be a reason for neglecting the activation entropy effects in previous studies of dislocation cross-slip~\cite{vegge_determination_2000}.

We now seek a close-form expression for $\Delta S_{\rm c}$ as a function of stress, which will not only reveal more insight on the physical nature of the activation entropy, but also provide a needed tool for predicting cross-slip rate in mesoscale models such as discrete dislocation dynamics~\cite{hussein_microstructurally_2015,longsworth_investigating_2021}.
We begin by defining $\tilde{\bm{\sigma}}$ as the stress of the crystal at zero temperature when subjected to the strain $\bm{\varepsilon}_T$, i.e. $\bm{\varepsilon}_T =\bm{\varepsilon}(\tilde{\bm{\sigma}}, 0) =\bm{\varepsilon}(\bm{\tau}_{\rm app}, T)$. 
%
$\tilde{\bm{\sigma}}$ is the stress in the simulation cell when performing MEP calculations for $E_{\rm c}(\bm{\varepsilon}_T)$; hence there is a one-to-one correspondence between $\bm{\varepsilon}_T$ and $\tilde{\bm{\sigma}}$.
At temperature $T$, the stress of the crystal subjected to strain $\bm{\varepsilon}_T$ is simply $\bm{\tau}_{\rm app}$.
But if the temperature is set to zero with the strain fixed at $\bm{\varepsilon}_T$, the stress value changes, i.e. $\tilde{\bm{\sigma}} = \bm{\tau}_{\rm app}+\hat{\sigma}\mathbf{I}+\bm{\tau}_{\rm ex}$, where $\hat{\sigma}$ is a hydrostatic (tensile) stress, and $\bm{\tau}_{\rm ex}$ is an excess shear stress.
We performed 500 MEP calculations of cross-slip at different stress $\tilde{\bm{\sigma}}$ and fit the activation energy $\tilde{H}_{\rm c}(\tilde{\bm{\sigma}}) = E_{\rm c}(\bm{\varepsilon}_T)$ results as a function of $\tilde{\bm{\sigma}}$ (see \hl{Supplementary Text IV}).
The functional form of $\tilde{H}_{\rm c}(\tilde{\bm{\sigma}})$ is a generalization of the $H_{\rm c}(\bm{\tau}_{\rm app})$ function established in our previous work~\cite{kuykendall_stress_2020}, and reduces to $H_{\rm c}(\bm{\tau}_{\rm app})$ when $\hat{\sigma}=0$.
Given the analytic function $\tilde{H}_{\rm c}(\tilde{\bm{\sigma}})$, we obtain the following expression for the activation entropy (see \hl{Supplementary Text V})

\begin{equation}
    \Delta S_{\rm c} = - K \alpha_V \, 
    \left(\frac{\partial \tilde{H}_{\rm c}}{\partial \hat{\sigma} }\right)
    + 
   \frac{1}{\mu}\left(\frac{\partial\mu}{\partial T}\right) \,
    \left(\frac{\partial \tilde{H}_{\rm c}}{\partial\bm{\tau}}\right)
    \cdot{\bm{\tau}}
    \label{eq:ScH_est_simplified}
\end{equation}
where $K$ is bulk modulus, $\alpha_{V}$ is volumetric thermal expansion coefficient, and $\mu$ is shear modulus.
Fig.~3 shows that Eq.~(\ref{eq:ScH_est_simplified}) agrees very well with the activation entropy computed above.
The two terms in Eq.~(\ref{eq:ScH_est_simplified}) can be identified as the contributions from thermal expansion and thermal softening effects to the activation entropy.
Both terms vanishes at the zero-stress limit.
Eq.~(\ref{eq:ScH_est_simplified}), combined with Eq.~(\ref{eq:rHTST_DS}), leads to a theoretical model that accurately predicts the cross-slip rate as a function of applied stress.
It can serve as an essential input for mesoscale models such as discrete dislocation dynamics~\cite{hussein_microstructurally_2015, akhondzadeh_statistical_2021} 
Because Eq.~(\ref{eq:ScH_est_simplified}) expresses $\Delta S_{\rm c}$ in terms of fundamental materials parameters and stress dependence of activation enthalpy, it is generally applicable to all stress-driven thermally activated processes in solids, such as phase transformation and twinning.

In conclusion, we have resolved a long-standing discrepancy between TST and direct MD predictions of cross-slip rate, and show that the anomalously large activation entropy is ultimately caused by the increasing shear and volumetric strain with increasing temperature at constant applied stress.
%
%
These anharmonic effects, i.e. thermal softening and thermal expansion, although previously ignored, can lead to orders-of-magnitude changes in the prediction of cross-slip rate.
%
%
%
%
We obtain an analytical expression for the activation entropy, which not only provides accurate predictions of cross-slip rate for meso-scale models, but also shows that our findings are generally applicable to all stress-driven thermally activated processes in solids.

\section*{Methods}

{\bf Prepare a single screw dislocation under applied stress.}
The dislocation structure is similar to our previous works \cite{kuykendall_stress_2020, kang_stress_2014}.
%
%
We start with a perfect fcc nickel crystal (lattice constant $a_0=\SI{3.52}{\angstrom}$) with simulation box dimension of $20[1\bar{1}0]\times20[111]\times10[\bar{1}\bar{1}2]$.
\SI{10}{\%} of the atoms are removed on each side of the $y$-direction to create free surfaces, resulting in with \num{78400} atoms in the simulation cell.
%
%
A single straight left-hand screw dislocation is created at the center of the $yz$-plane with Burger's vector $\mathbf{b}=a_0[\bar{1}10]/2$ along the positive $x$-direction $\bm{\xi}=[1\bar{1}0]$.
The initial configuration is obtained by splitting the screw dislocation into two Shockley partial dislocations (orange arrows in Fig.~1(a)) with stacking fault on the gliding plane, i.e., the $(111)$ plane~\cite{kuykendall_stress_2020}.

We perform energy minimization with applied shear stresses 
$\bm{\tau}_{\rm app} = (\sigma_{\rm e}^{\rm g}, \sigma_{\rm s}^{\rm c}, \sigma_{\rm e}^{\rm c})$
to the dislocation structure.
The Cartesian stress tensor can be calculated from the applied stress as,
%
%
\begin{align}
    \sigma_{\rm s}^{\rm g} = \sigma_{xy},\quad
    \sigma_{\rm e}^{\rm g} = \sigma_{yz},\quad
    \sigma_{\rm s}^{\rm c} = \frac{2\sqrt{2}\sigma_{xz}-\sigma_{xy}}{3}, \quad
    \sigma_{\rm e}^{\rm c} = \frac{7\sigma_{yz}+2\sqrt{2}\left(\sigma_{zz}-\sigma_{yy}\right)}{9} \label{eq:sigescart}
\end{align}
\noindent where $\sigma_{zz} = -\sigma_{yy}$ is enforced to enable a one-on-one mapping between the Cartesian stress and the Escaig-Schmid stress components, and $\sigma_{xy}$ is set to be zero to avoid the screw dislocation moving on the original slip plane.

On the one hand, due to free surfaces in the $y$-direction, 
stress components $(\sigma_{xy}, \sigma_{yy}, \sigma_{yz})$ are applied by external forces $\mathbf{f}_y=({A}/{N_{xz}})(\sigma_{xy}, \sigma_{yy}, \sigma_{yz})$ to the first layer of atoms ($N_{xz}=1600$ atoms in total) on the free surfaces,
and $A=H_xH_z$ is the area of the surface.
On the other hand, due to the periodic boundary condition on $x$- and $z$- directions, the stress components $(\sigma_{xx}, \sigma_{zz}, \sigma_{xz})$ are controlled by adjusting the components $(H_x, H_z, H_{xz})$ in the simulation cell iteratively until the stresses are converged.
The simulation cell matrix (cell vectors) $\mathbf{H} = \left[\mathbf{c}_1|\mathbf{c}_2|\mathbf{c}_3\right]$ is defined as,
\begin{align}
    \mathbf{H} =
    \begin{bmatrix}
        H_x & H_{xy} & H_{xz}\\
        0 & H_y & H_{yz}\\
        0 & 0 & H_z\\
    \end{bmatrix}
\end{align}
%
%
%

The stress of the dislocation configuration is calculated by averaging the atomic stress~\cite{thompson_general_2009} of all the atoms \SI{20}{\angstrom} below the free surfaces to avoid the surface effect.
The convergence tolerance of the stress is $\pm\SI{0.05}{MPa}$.

{\bf Minimum-energy path (MEP) search.}
To perform MEP search, we first prepare the initial state $A$ before the transition state and the final state $B$ after the transition state.
The converged metastable dislocation structure from the previous section is used as the initial state $A$.
The final state $B$ is prepared with the same full screw dislocation structure as state $A$, but with the middle half of the dislocation dissociated on the cross-slip plane $(11\bar{1})$, while the rest of the dislocation still dissociates on $(111)$~\cite{kuykendall_stress_2020}.
Energy minimization is then performed to obtain the final state $B$ under the same applied shear stress $\bm{\tau}_{\rm app}$.
In order to obtain a better initial guess and help with the convergence of the MEP search, the conjugate-gradient energy minimization on the final state $B$ is only performed for five iterations so that the cross-slipped dislocation does not move towards the free surface and annihilate, i.e., the state $B$ is not too far away from the transition state.
%
%
Starting from a linear interpolation (\num{32} image copies) between states $A$ and $B$ as the initial guess, the MEP search is performed using the free-end string method~\cite{e_string_2002} with reparameterization and trimming~\cite{kuykendall_investigating_2015}.
After the string method is converged, we use the dimer method~\cite{henkelman_dimer_1999} to obtain the exact transition state $S$.
Starting from the two images closest to the maximum value as the initial dimer, we iteratively shrink the the dimer until the distance is below \SI{e-7}{\angstrom}.
The external forces $\mathbf{f}_y$ and simulation cell matrix $\mathbf{H}$ from state $A$ are applied during all the energy minimization steps in state-$B$ preparation, MEP search, and dimer method to ensure the same applied stress condition $\bm{\tau}_{\rm app}$.

{\bf Molecular dynamics (MD) simulation.}
MD simulations of dislocation cross-slip are performed using the LAMMPS package \cite{thompson_lammps_2022}.
To prepare the dislocation structure at finite temperature $T$ under the applied stress condition $\bm{\tau}_{\rm app}$, we start from the state $A$ with zero applied stress.
The system is gradually heated up to the target temperature $T$ and  equilibrated for \SI{10}{ps} using the Nos\'e-Hoover thermostat~\cite{nose_molecular_2002} with zero stress applied, to avoid premature cross-slip.
The configuration is then gradually loaded to the target stress $\mathbf{\tau}_{\rm es}$ and further equilibrated for \SI{2}{ps}.
The method to control the stress is the same as in the previous sections.
After the system is equilibrated, we apply a small random perturbation (uniform distribution with the magnitude of \SI{\pm e-4}{\angstrom\cdot{s^{-1}}}) to the initial velocity before continuing the MD simulation to avoid repeated MD trajectories.
The MD simulation is continued until cross slip occurs (sudden release of the applied stress) and the cross-slip time $t_{\rm cs}$ is recorded.

{\bf Harmonic vibrational frequencies.}
The product of the harmonic vibrational frequencies in Eq.~(\ref{eq:rHTST}) is obtained from the Hessian matrices of the initial state $A$ ($\mathbf{K}_A$) and the transition state $S$ ($\mathbf{K}_S$).
The standard approach to obtain the prefactor is to diagonalize $\mathbf{K}_A$ and $\mathbf{K}_S$.
However, for our system ($N=\num{78400}$), the Hessian matrices have a size of $3N\times3N = \num{235200}\times\num{235200}$, which requires significant computational load.
Instead, we can calculate the products of the eigenfrequencies from the determinant if and only if $\mathbf{K}$ is \emph{non-singular}.

To avoid the non-sigularity from the three rigid-body translational modes (eigen frequency $\nu=0$), we couple three soft harmonic spring forces $k$ to the $x$, $y$, and $z$-directions on one atom (atom \#1) in both states $A$ and $S$.
This is equivalent to modifying the first three diagonal elements of the Hessian matrix $\mathbf{K}$,
\begin{align}
    K_{11} \rightarrow K_{11} + k;\quad
    K_{22} \rightarrow K_{22} + k;\quad
    K_{33} \rightarrow K_{33} + k
    \label{eq:modified_Hessian}
\end{align}

We then obtain the product of the eigen frequencies by calculating the determinant of the modified Hessian matrix using the sparse LU decomposition in MATLAB.
The negative eigenvalue of the Hessian matrix at state $S$ is obtained by finding the minimum eigenvalue using the `eigs' method in MATLAB.
The soft spring frequencies are selected to be $k=\SI{1e-4}{eV/\angstrom}$, which will be cancelled out while calculating the prefactor $\nu_{\rm HTST}$ from taking the ratio between the determinant of state $S$ and state $A$.
A detailed proof of the method is provided in \hl{Supplementary Text VI}.

\bibliography{references}

\clearpage

\begin{figure}[!hbt]
    \centering
    \includegraphics[width=0.45\linewidth]{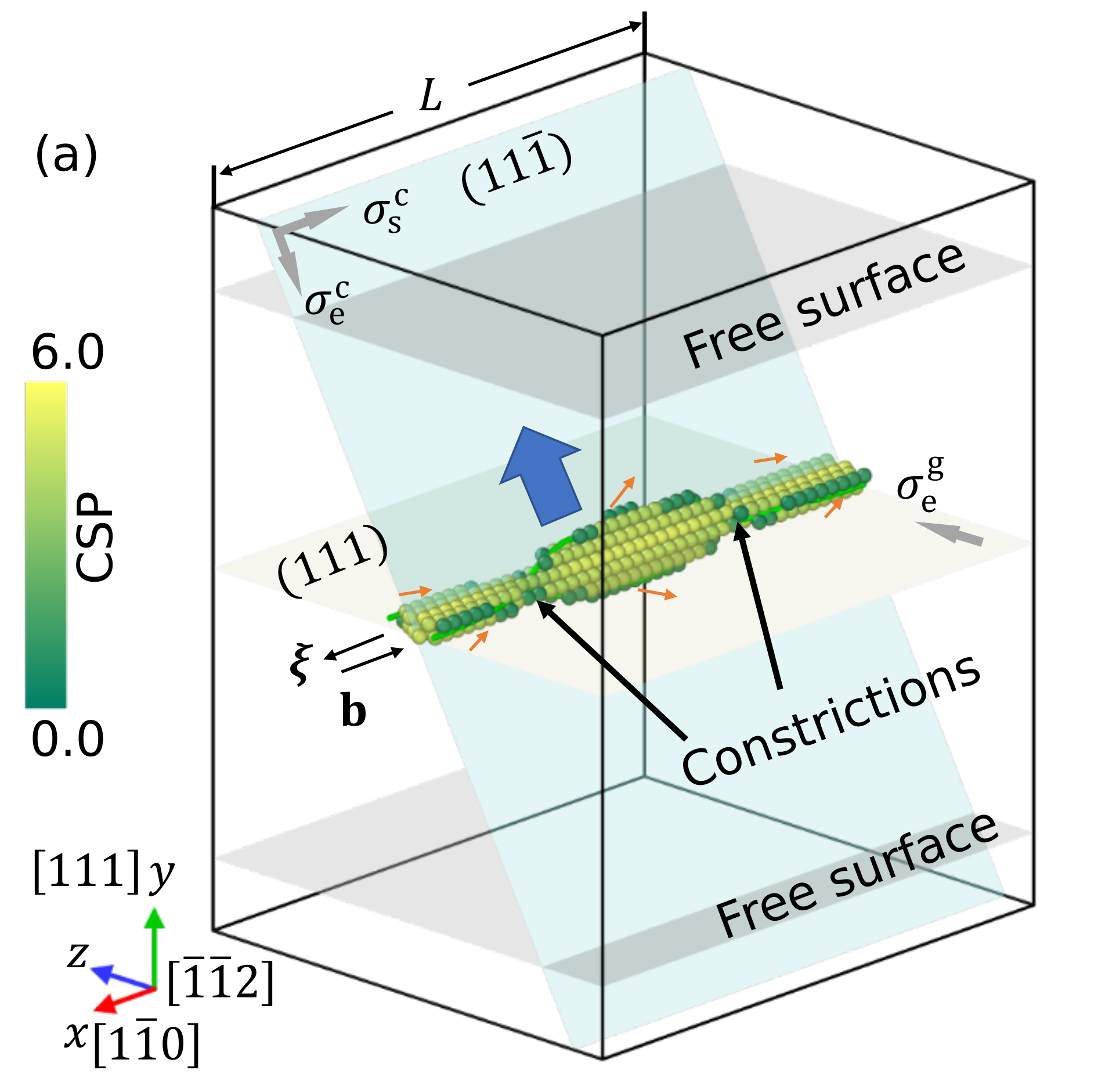}
    \includegraphics[width=0.9\linewidth]{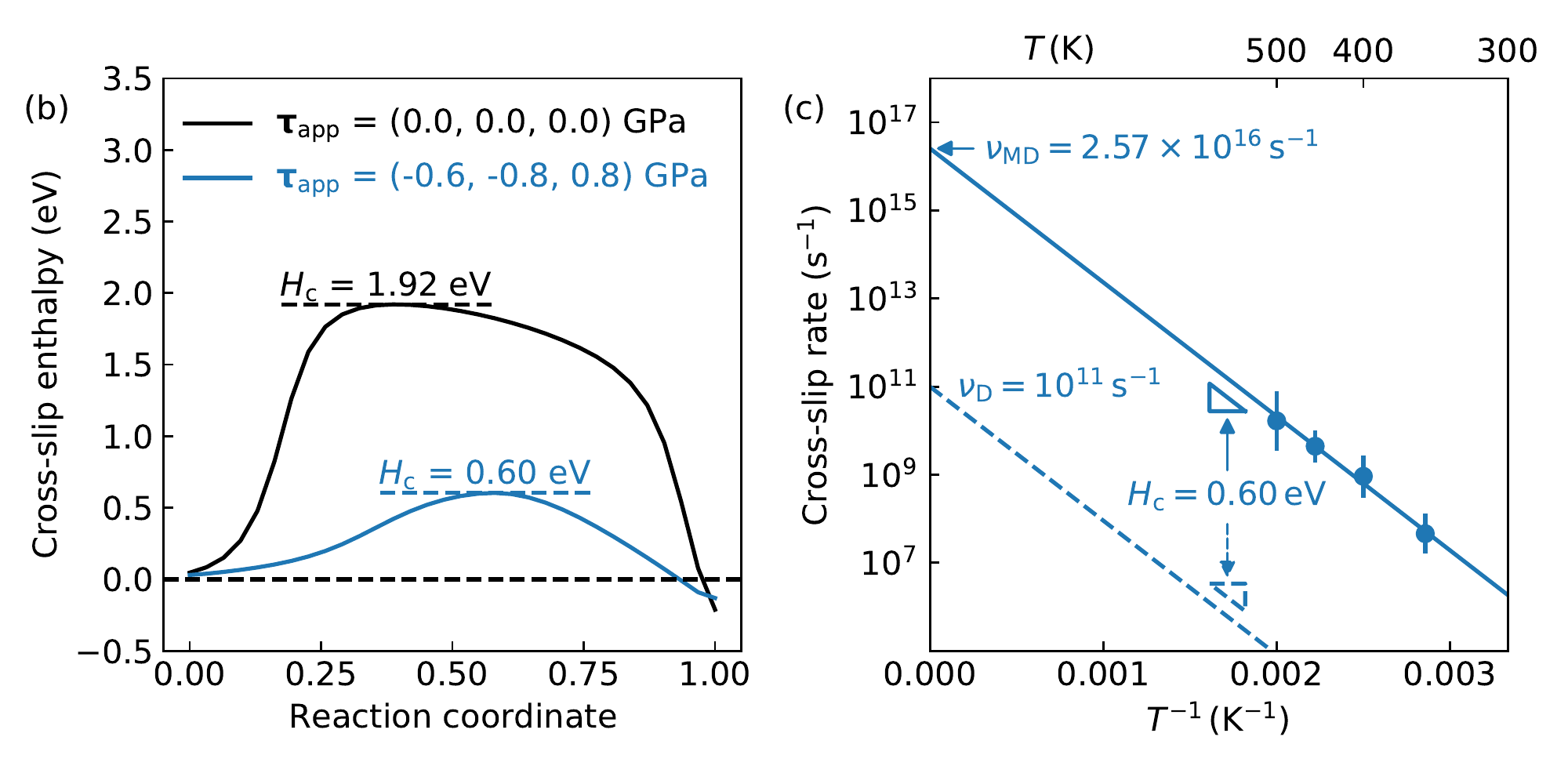}
\end{figure}

\noindent {\bf Fig.~1. Discrepancy between MD rates and TST predictions}.
(a) Simulation cell ($20[1\bar{1}0]\times20[111]\times10[\bar{1}\bar{1}2]$) with a screw dislocation along the $x$-direction, visualized by OVITO \cite{stukowski_visualization_2009}.
Atoms are colored according to their centrosymmetric parameter (CSP), given an fcc crystal structure (12 neighbors).
The atoms with ${\rm CSP} > 1$ are extracted to visualize the dislocation core structure.
The screw dislocation changes the slip plane from $(111)$ to $(11\bar{1})$ following the Freidel-Escaig mechanism \cite{puschl_models_2002}.
The three Escaig-Schmid stresses components $\bm{\tau}_{\rm app} = (\sigma_{\rm e}^{\rm g}, \sigma_{\rm s}^{\rm c}, \sigma_{\rm e}^{\rm c})$ controlling the cross-slip process are applied on the two slip planes.
The cross-slipped dislocation moves along the cross-slip plane (blue arrow) if $\sigma_{\rm s}^{\rm c}$ is applied and finally annihilates at the free surface.
(b) Converged minimum-energy paths of cross-slip, calculated at zero stress $\bm{\tau}_{\rm app} = 0$ and fixed applied stress $\bm{\tau}_{\rm app} = (-0.6,-0.8, 0.8)\,{\rm GPa}$.
The positive directions of the shear stresses are marked as arrows in figure (a).
(c) Cross-slip rates at different temperatures obtained by MD simulations (solid line) and predicted by TST Eq.~(\ref{eq:Arrhenius_rate}) (dashed line) under fixed applied stress $\bm{\tau}_{\rm app}$.

\clearpage

\begin{figure}[!hbt]
    \centering
    \includegraphics[width=0.9\linewidth]{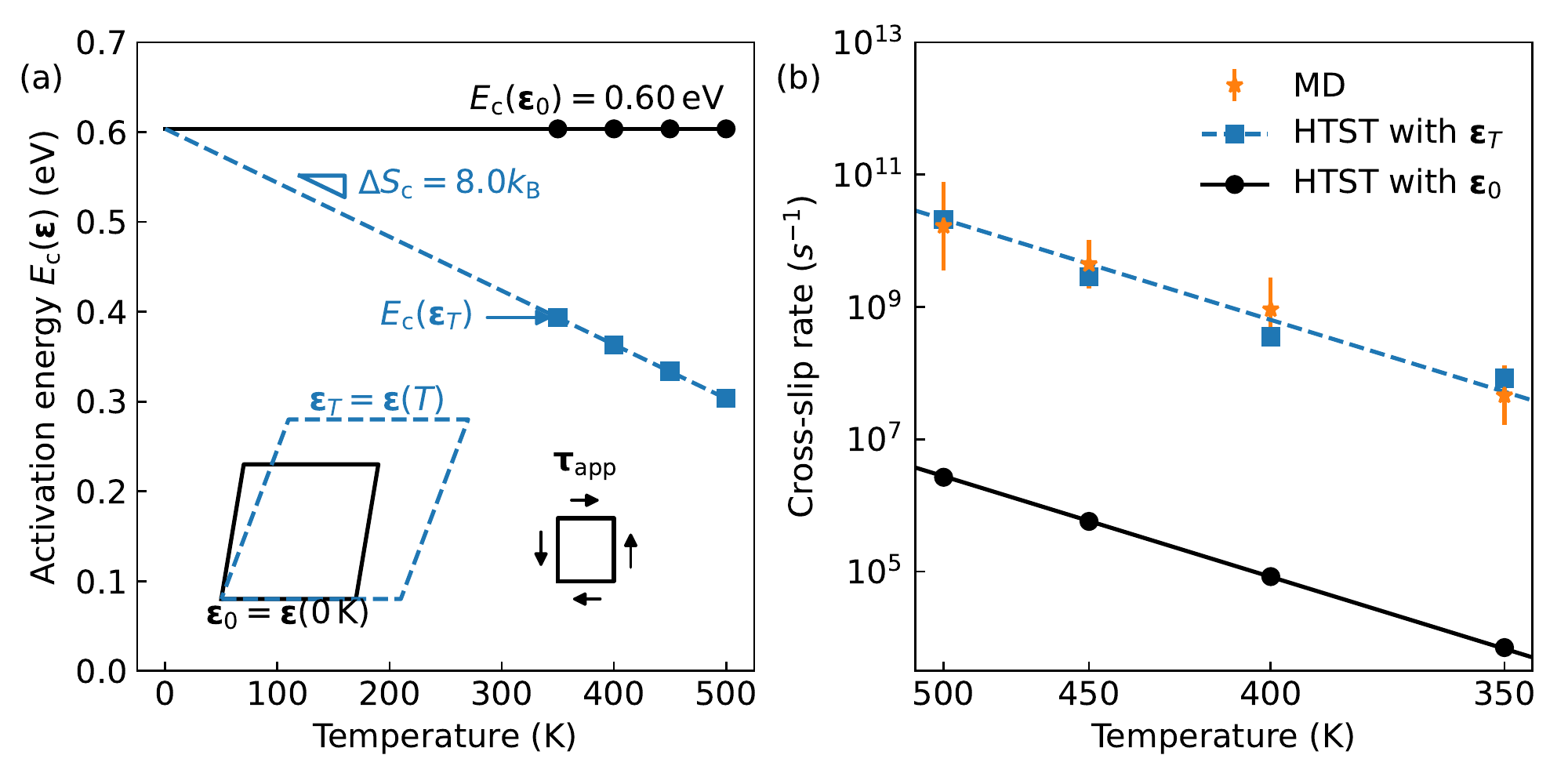}
\end{figure}

\noindent {\bf Fig.~2. Activation entropy due to the thermal strain}.
(a) Activated energy calculated at zero-temperature strain $\bm{\varepsilon}_0$ and corresponding finite-temperature strain $\bm{\varepsilon}_{\rm T}$.
The inset diagram schematically shows the thermal strain caused by temperature increase with the same applied stress $\bm{\tau}_{\rm app}$.
(b) Estimated rates using HTST (Eq.~(\ref{eq:rHTST})) with the activation energy and prefactor evaluated at $\bm{\varepsilon}_0$ and $\bm{\varepsilon}_{\rm T}$.
The benchmark MD rates are shown as the stars with the error bar.

\clearpage

\begin{figure}[!hbt]
    \centering
    \includegraphics[width=0.6\linewidth]{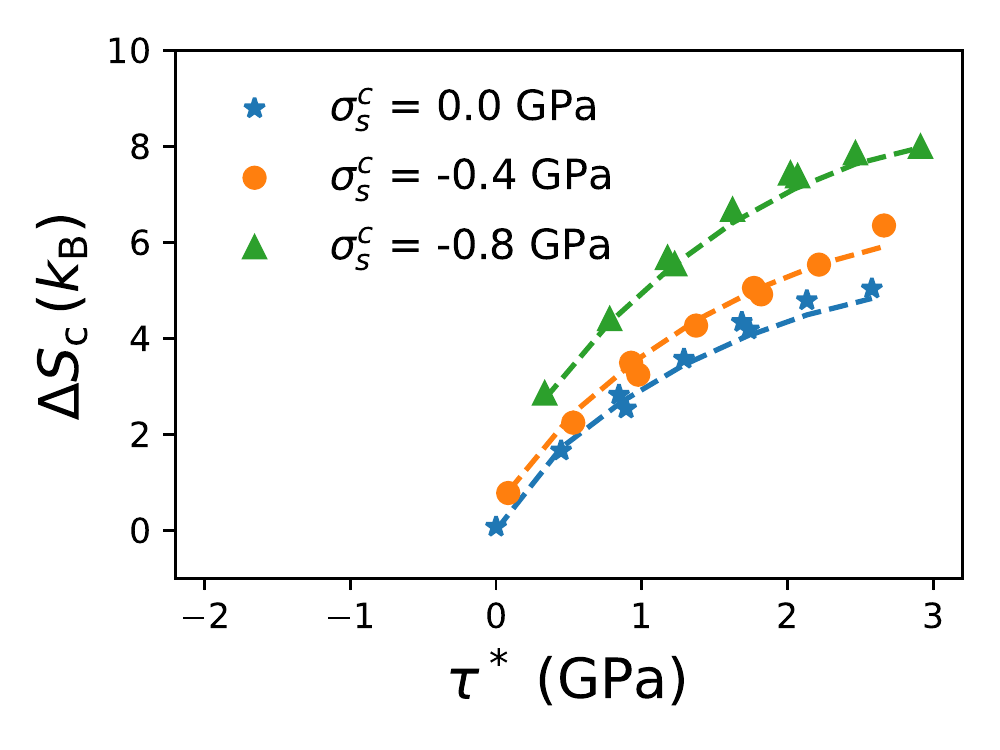}
\end{figure}

\noindent {\bf Fig.~3. Activation entropy}.
%
%
Solid dots represent the activation entropy of \num{27} applied Escaig-Schmid stress conditions from MEP calculations using the same method as Fig.~2(a).
The dashed lines are the estimated activation entropy from Eq.~(\ref{eq:ScH_est_simplified}).

\clearpage

\appendix

\section*{Supplementary Text}

\subsection*{I. Correction to the soft vibrational modes}

Fig.~S1(a) shows the Goldstone mode, in which the constriction can move along the dislocation line direction.
To obtain the energy profile along the Goldstone mode, we start from the transition state $S$.
We move all the atoms along the line direction by one Burger's vector \hl{to obtain the next repeated configuration} (local minimums in Fig.~S1).
The energy profile of Goldstone mode is obtained by linear interpolation between the repeated transition states, shown as a solid line in Fig.~S1(b), with the harmonic approximation of the basin illustrated as the dashed line.
Unlike in reference~\cite{proville_quantum_2012}, since the applied stress in our work is not very high, the Goldstone mode (constrictions) at the activated state do not move freely along the dislocation line due to the lattice frictions.
It can be seen that the actual energy profile is significantly deviating from the harmonic approximation, which causes a massive error to the partition function and the free energy.
Fig.~S1(b) only shows one period of $E_{\rm S}^{\rm G}$, since the lattice in the $[1\bar{1}0]$ direction is repeated for $N_{\rm cs}=L_x/b$ times in the dislocation line direction.
The correcting prefactor to this soft vibrational mode can be written as the ratio between the partition functions of the 1D energy profile ($^{\rm G}$) and the harmonic approximation ($^{\rm H}$),

\begin{align}
    \tilde{\nu}_{\rm S} = \frac{z_{\rm S}^{\rm H}}{z_{\rm S}^{\rm G}} &= 
    \frac{\int_{-\infty}^{\infty}\,\exp[-E_{\rm S}^{\rm H}(x)/k_{\rm B}T]\,{\rm d}x}
         {N_{\rm cs}\int_{-R/2}^{R/2}\,\exp[-E_{\rm S}^{\rm G}(x)/k_{\rm B}T]\,{\rm d}x} \nonumber \\
    &= \frac{\sqrt{Kb^2/2\pi k_{\rm B}T}}
         {L_x\int_{-R/2}^{R/2}\,\exp[-E_{\rm S}^{\rm G}(x)/k_{\rm B}T]\,{\rm d}x}
\end{align}
where $K$ is the curvature of the harmonic approximation at $x=0$, indicated by the dashed quadratic curve in Fig.~S1(b), and $R$ is the atom displacement of one period.
The denominator is evaluated by numerical integration of the energy profile $E_{\rm S}^{\rm G}(r)$ illustrated as the solid curve.
The correction factor can be as high as $\tilde{\nu}_S\sim7.0$, contributing about one order of magnitude to the rate prediction.

Similarly, the gliding mode of the initial state is shown in Fig.~S2. We move all the atoms along the gliding direction by one \hl{gliding vector $\frac{a}{6}[\bar{1}\bar{1}2]$ to obtain the next repeated configuration}. The energy profile of gliding is then obtained by MEP search.
It is readily seen that the energy barrier is much higher than the Goldstone mode, and the harmonic approximation is working better estimating the partition function compared to the Goldstone mode.

\begin{align}
    \tilde{\nu}_{\rm A} &= \frac{z_{\rm A}^{\rm H}}{z_{\rm A}^{\rm G}} = 
    \frac{\int_{-\infty}^{\infty}\,\exp[-E_{\rm A}^{\rm H}(z)/k_{\rm B}T]\,{\rm d}z}
         {\int_{-R/2}^{R/2}\,\exp[-E_{\rm A}^{\rm G}(z)/k_{\rm B}T]\,{\rm d}z} \nonumber \\
    &= \frac{\sqrt{K_Ab^2/2\pi k_{\rm B}T}}
         {\int_{-R/2}^{R/2}\,\exp[-E_{\rm A}^{\rm G}(z)/k_{\rm B}T]\,{\rm d}z}
\end{align}
where $K_A$ is the curvature of the harmonic approximation.
The correction factor $\tilde{\nu}_A$ is around $\sim1.4$, much smaller compared to the Goldstone mode.
It is worth noting that the gliding mode does not need to multiply by the number of repeat lattices in the $z$-direction,
since the cross-slip only occur at the intersection between the glide plane and the cross-slip plane.
All the other positions of the dislocation are not contributing to the partition function.

These soft modes are accounted for by numerically evaluating the partition function in these eigen directions.
The dislocation length $L$ is embedded in $\tilde{\nu}_{\rm S}$ since the repeated lattice number of the Goldstone mode is $N_{\rm G} = L/b$. 
The effect of these soft vibrational modes can be significant at the applied stress condition 
($\tilde{\nu}_{\rm A}/\tilde{\nu}_{\rm S}\approx0.2$).
We believe that these two terms in the prefactor $\nu_0(\bm{\varepsilon})=
\nu_{\rm HTST}\,{\tilde{\nu}_{\rm A}}/{\tilde{\nu}_{\rm S}}$
cover all the important vibrational contribution to the free energy difference 
$F_{\rm c}(\bm{\varepsilon})$.

\subsection*{II. Additional calculations for different applied stresses}

To assess the generality of the conclusions, we repeat the MD simulations and MEP calculations for cross-slip at two different applied stress conditions of $\bm{\tau}_{\rm app}=(-0.8,-0.8,0.8)\si{GPa}$ and $\bm{\tau}_{\rm app}=(-0.8,-0.8,0.6)\si{GPa}$, as shown in Fig.~S3.
After using the finite-temperature strain $\bm{\varepsilon}_T$ to calculate the activation energy, the HTST estimate (blue squares) of the cross-slip rate matches with the MD results (orange stars) for different applied stress conditions.
This result indicates that the thermal expansion and the thermal softening effects contribute primarily to the large activation entropy at a given applied stress.
The activation entropy $\Delta S_{\rm c}$ is evaluated as the negative slope of the activation energy decrease with increasing temperature.
%

To examine the stress-dependence of the activation entropy $\Delta S_{\rm c}$ described in Eq.~(\ref{eq:cross-slip:ScH_est}),
we perform the MEP calculations of \num{27} applied stress conditions,
where the three shear stress components $-\sigma_{\rm e}^{\rm g}$, $-\sigma_{\rm s}^{\rm c}$, and $\sigma_{\rm e}^{\rm c}$ varies among \num{0.0}, \num{0.4}, \SI{0.8}{GPa}.
The calculated activation entropy is marked as solid markers in Fig.~3, as a function of the effective shear stress $\tau^*$ from Eq.~(\ref{eq:cross-slip:effective_tau}).
It is readily seen that the activation entropy can be grouped into three curves by the value of $\sigma_{\rm s}^{\rm c}$.
%
%
These stress-dependent activation entropy values are then modeled by Eq.~(\ref{eq:ScH_est_simplified}) (\hl{Supplementary Text V}) and shown as dashed lines in Fig.~3.

\subsection*{III. Constant strain activation entropy from HTST}

Based on TST, the cross-slip rate can be written as a function of constant strain $\bm{\varepsilon}$ and temperature $T$~\cite{ryu_predicting_2011},

\begin{align}
    r_{\rm TST}(\bm{\varepsilon},T)&= 
        \frac{k_{\rm B}T}{h}\,
        \exp\left[-\frac{F_{\rm c}(\bm{\varepsilon},T)}{k_{\rm B}T}\right] \nonumber\\
        &=
        \frac{k_{\rm B}T}{h}\,
        \exp\left[\frac{S_{\rm c}(\bm{\varepsilon})}{k_{\rm B}}\right]
        \exp\left[-\frac{E_{\rm c}(\bm{\varepsilon})}{k_{\rm B}T}\right]
    \label{eq:adx:rTST}
\end{align}
\noindent where $G_{\rm c} = G_S - G_A$ is the activation Gibbs free energy, the Gibbs free energy difference between the transition state $S$ and initial state $A$.
The activation Gibbs free energy can be expressed as $G_{\rm c}(\bm{\varepsilon},T) = H_{\rm c}(\bm{\varepsilon}) - TS_{\rm c}(\bm{\varepsilon})$, where $H_{\rm c}$ and $S_{\rm c}$ are the activation enthalpy and activation entropy depending on the strain $\bm{\varepsilon}$, respectively.
Combining Eq.~(\ref{eq:rHTST}) and Eq.~(\ref{eq:adx:rTST}), the activation entropy can be obtained from HTST calculation,
\begin{align}
    S_{\rm c} = k_{\rm B}\ln\left(
        \nu_{\rm HTST}\,\frac{\tilde{\nu}_{\rm A}}{\tilde{\nu}_{\rm S}}\frac{h}{k_{\rm B}T}
        \right)
\end{align}

\subsection*{IV. Analytical expressions for stress-dependent activation enthalpy $H_{\rm c}$}

As we discussed in the main text (Fig.~3), the activation energy at finite temperature strain is equivalent to the activation enthalpy at zero temperature with excess stress applied.
Since the MEP search algorithm is computationally expensive, it is not possible to calculate activation enthalpy everytime with a new applied stress condition.
In this section, we will build an analytical formula for the activation enthalpy as a function of applied stress.
Kuykendall et al.~\cite{kuykendall_stress_2020} and Esteban-Manzanares et al.~\cite{esteban-manzanares_influence_2020} studied the zero-temperature stress-dependence of the cross-slip activation enthalpy, but the isotropic stress $-\hat{\sigma}\mathbf{I}$ is not considered in both works.
Here we develop an $H_{\rm c}$ function based on Kuykendall et al. (2020)'s \cite{kuykendall_stress_2020} expression, considering the applied Escaig-Schmid stress components
$\bm{\tau}_{\rm app} = (\sigma_{\rm e}^{\rm g}, 
                        \sigma_{\rm s}^{\rm c},
                        \sigma_{\rm e}^{\rm c})$
and the isotropic stress $\hat{\sigma}$:

\begin{align}
    H_{\rm c}(\tau^*,\hat{\tau}_0)=
    A\left[1-\left(\frac{\tau^*}{\hat{\tau}_0}\right)^p\right]^q
    \label{eq:cross-slip:activation_enthalpy}
\end{align}
\noindent where the effective shear stress $\tau^*$ and the effective cross-slip stress $\hat{\tau}_0$ (where the energy barrier is zero) are defined as,

\begin{align}
    &\tau^*(\sigma_{\rm e}^{\rm g}, 
           \sigma_{\rm e}^{\rm c},
           \sigma_{\rm s}^{\rm c}) 
           = C_{\rm e}^{\rm g}\sigma_{\rm e}^{\rm g} 
            + C_{\rm e}^{\rm c}\sigma_{\rm e}^{\rm c} 
            +(D_{\rm s}^{\rm c}\sigma_{\rm s}^{\rm c})^2
            \label{eq:cross-slip:effective_tau} \\
    &\hat{\tau}_0(-\hat{\sigma},\sigma_{\rm s}^{\rm c}) 
            = \tau_0 - K_1(-\hat{\sigma})(\sigma_{\rm s}^{\rm c})^2
             - K_2(-\hat{\sigma})^{K_3}
            \label{eq:cross-slip:effective_Peierl} 
\end{align}

We fit this expression based on 500($5\times5\times5\times4$) MEP calculations, with $-\sigma_{\rm e}^{\rm g}$, $-\sigma_{\rm s}^{\rm c}$, and $\sigma_{\rm e}^{\rm c}$ varies among \num{0.0}, \num{0.2}, \num{0.4}, \num{0.6}, and \SI{0.8}{GPa},
and the hydrostatic stress $-\hat{\sigma}$ varies among \num{0.0}, \num{2.0}, \num{4.0}, and \SI{6.0}{GPa}.
In these simulations, $\sigma_{\rm e}^{\rm g}$ is negative (when it is non-zero), to promote constriction of the stacking fault,
while $\sigma_{\rm e}^{\rm c}$ is positive (when it is non-zero) to promote expansion of the stacking fault on the cross-slip plane. 
In our simulations, $\sigma_{\rm s}^{\rm c}$ is negative (when it is non-zero), though both positive and negative Schmid stresses on the cross-slip plane are expected to promote cross-slip.
Each MEP relaxation is performed for \num{800} iterations (which takes \SI{8}{hours} on \num{32} cores) and is considered converged if
(1) the slope of the linear fit of energy barrier over the last 400 steps is less than \SI{e-5}{eV/step};
and (2) the mean squared error of the linear fit is less than \SI{e-3}{eV^2}.
If convergence is not reached, the MEP relaxation is restarted for another 400 iterations.
If the energy barrier during relaxation becomes negative, we consider the calculation has failed to converge.
All these \num{500} simulations (whose initial paths are constructed based on the FE mechanism) converge to MEPs that are consistent with the FE mechanism.
\begin{table}[!ht]
    \centering
    \begin{tabular}{c|c}
        $C_{\rm e}^{\rm g}$ & -2.1077 \\
        $C_{\rm e}^{\rm c}$ & 1.1150 \\
        $D_{\rm s}^{\rm c}$ & 0.7218 \\
        \hline
        $A$ & 1.9193 \\
        $p$ & 0.7711 \\
        $q$ & 1.4428 \\
        $\tau_0$ & 5.5949 \\
        \hline
        $K_1$ & 0.1592 \\
        $K_2$ & 0.3620 \\
        $K_3$ & 0.6994
    \end{tabular}
    \caption{Fitting parameters for Equation~(\ref{eq:cross-slip:activation_enthalpy}),
            (\ref{eq:cross-slip:effective_tau}),(\ref{eq:cross-slip:effective_Peierl})
    }
    \label{tab:cross-slip:fitting_parameters}
\end{table}
The fitted numerical values for the seven fitting parameters $(A, p, q, \tau_0, C_{\rm e}^{\rm g}, C_{\rm e}^{\rm c}, D_{\rm s}^{\rm c})$ are given in 
Table~\ref{tab:cross-slip:fitting_parameters}.
It is worth noting that except for the magnitude $A$ due to the smaller simulation cell in this work, the values of the parameters are the same as 
Kuykendall et al. (2020) \cite{kuykendall_stress_2020},
indicating the consistency of this analytical expression.
Fig.~S4 shows the fitted activation enthalpy as a function of $\tau^*$ and $\hat{\tau}_0$.
The formula accurately predicts the activation enthalpy at any given applied stress $\tilde{\bm{\sigma}}$ within the fitting range.

\subsection*{V. Analytical expressions for stress-dependent activation entropy $S_{\rm c}$}

The activation entropy $\Delta S_{\rm c}$ is defined as the reduction of activation energy with increasing temperature and fixed stress condition $\bm{\sigma} = \bm{\tau}_{\rm app}$.
Using Legendre transformation, the activation energy as a function of strain can be written in terms of activation enthalpy as a function of stress, i.e., $E_{\rm c}(\bm{\varepsilon}_T) = H_{\rm c}(\tilde{\bm{\sigma}})$.
Here the $\tilde{\bm{\sigma}}$ is the corresponding stress for the finite-temperature strain $\bm{\varepsilon}_T$ at \SI{0}{K}, i.e., $\tilde{\bm{\sigma}} = \bm{\sigma} (\bm{\varepsilon}_T, \SI{0}{K})$, where $\bm{\varepsilon}_T = \bm{\varepsilon} (\bm{\tau}_{\rm app}, T)$ is the finite-temperature strain.
At \SI{0}{K}, excess stresses need to be applied for the system to remain the strain $\bm{\varepsilon}_T$, as shown in Fig.~\hl{S5}(b)(c),

\begin{align}
    \tilde{\bm{\sigma}} = \hat{\sigma}\mathbf{I} + \bm{\tau} = \bm{\tau}_{\rm app} + \hat{\sigma}\mathbf{I} + \bm{\tau}_{\rm ex}
    \label{eq:excess_stress}
\end{align}
\noindent where $\hat{\sigma}\mathbf{I}$ is the excess isotropic stress and $\bm{\tau}_{\rm ex}$ is the excess shear stress.
As a result, the activation entropy can be decomposed as,

\begin{align}
    \Delta S_{\rm c} 
        &=
         \left(\frac{\partial E_{\rm c}(\bm{\varepsilon}_T)}{\partial T}\right)_{\bm{\sigma}=\bm{\tau}_{\rm app}}
         =\left(\frac{\partial H_{\rm c}(\tilde{\bm{\sigma}})}{\partial T}\right)_{\bm{\tau}_{\rm app}} \nonumber \\
         &=\left(\frac{\partial H_{\rm c}}{\partial\tilde{\bm{\sigma}}}\right)
         \left(\frac{\partial\tilde{\bm{\sigma}}}{\partial\bm{\varepsilon}_{T}}\right)
         \left(\frac{\partial\bm{\varepsilon}_{T}}{\partial T}\right)_{\bm{\tau}_{\rm app}}
    \label{eq:cross-slip:ScH_est}
\end{align}
where the first term $(\partial H_{\rm c}/\partial\tilde{\bm{\sigma}})$ is the stress gradient of the activation enthalpy, which can be calculated from the analytical expression Eq.~(\ref{eq:cross-slip:activation_enthalpy})~(\ref{eq:cross-slip:effective_tau})~(\ref{eq:cross-slip:effective_Peierl}) in the previous section.
The second term $(\partial\tilde{\bm{\sigma}}/\partial\bm{\varepsilon}_T)$ comes from the constitutive relationship at \SI{0}{K}.
The last term $(\partial\bm{\varepsilon}_{T}/\partial T)_{\bm{\sigma}=\bm{\tau}_{\rm app}}$ is the strain change with increasing temperature given the system remains at the applied stress $\bm{\tau}_{\rm app}$, as shown in Fig.~S5(b)(c).
This entropic effect can be further decomposed according to the isotropic component and the shear component of the stress,

\begin{align}
    \Delta S_{\rm c} &= 
        \left[
            \left(\frac{\partial H_{\rm c}}{\partial(-\hat{\sigma})}\right)
            \left(\frac{\partial(-\hat{\sigma})}{\partial\bm{\varepsilon}_T}\right) + 
            \left(\frac{\partial H_{\rm c}}{\partial\bm{\tau}}\right)
            \left(\frac{\partial\bm{\tau}}{\partial\bm{\varepsilon}_T}\right)
        \right]
        \left(
            \frac{\partial\bm{\varepsilon}_T}{\partial T}
        \right)_{\bm{\tau}_{\rm app}} \nonumber \\
    & = \left(\frac{\partial H_{\rm c}}{\partial(-\hat{\sigma})}\right)
        \left[
        \left(\frac{\partial(-\hat{\sigma})}{\partial\bm{\varepsilon}_T}\right)
        \left(\frac{\partial\bm{\varepsilon}_T}{\partial T}\right)_{\bm{\tau}_{\rm app}}
        \right] +
        \left(\frac{\partial H_{\rm c}}{\partial\bm{\tau}}\right)
        \left[
        \left(\frac{\partial\bm{\tau}}{\partial\bm{\varepsilon}_T}\right)
        \left(\frac{\partial\bm{\varepsilon}_T}{\partial T}\right)_{\bm{\tau}_{\rm app}}
        \right]
    \label{eq:cross-slip:ScH_decompose}
\end{align}

With the analytical expression Eq.~(\ref{eq:cross-slip:activation_enthalpy})~(\ref{eq:cross-slip:effective_tau})~(\ref{eq:cross-slip:effective_Peierl}), 
${\partial H_{\rm c}}/{\partial(-\hat{\sigma})}$ and ${\partial H_{\rm c}}/{\partial\bm{\tau}}$ can be evaluated analytically at any stress state.
Here we estimate the rest of the expression based on the following assumptions:
First, the isotropic component refers to the thermal expansion effect~\cite{mishin_calculation_2001},

\begin{align}
    \left[
    \left(\frac{\partial(-\hat{\sigma})}{\partial\bm{\varepsilon}_T}\right)
    \left(\frac{\partial\bm{\varepsilon}_T}{\partial T}\right)_{\bm{\tau}_{\rm app}}
    \right]
    =\frac{\partial P}{\partial T}
    =V\left(\frac{\partial P}{\partial V}\right)\cdot
     \frac{1}{V}\left(\frac{\partial V}{\partial T}\right)
    =K\alpha_V
    \label{eq:ScH_isotropic}
\end{align}
where $P$ is the hydrostatic pressure; $\alpha_V$ is the volumetric thermal expansion coefficient; and $K$ is the bulk modulus.
Second, the shear component expresses the thermal softening effect due to the decreasing shear modulus with increasing temperature~\cite{ryu_predicting_2011},
%

\begin{align}
    \left[
    \left(\frac{\partial\bm{\tau}}{\partial\bm{\varepsilon}_T}\right)
    \left(\frac{\partial\bm{\varepsilon}_T}{\partial T}\right)_{\bm{\tau}_{\rm app}}
    \right]
    \approx \frac{1}{\mu}\left(\frac{\partial\mu}{\partial T}\right) 
    \cdot\bm{\tau} 
    \label{eq:ScH_shear}
\end{align}
where $\mu$ is the shear modulus, and $(\partial\mu/\partial T)$ is the gradient of shear modulus respect to temperature.
Combining Eq.~(\ref{eq:ScH_isotropic}) and (\ref{eq:ScH_shear}) into Eq.~(\ref{eq:cross-slip:ScH_decompose}), we reach the analytical model for predicting stress-dependent activation entropy Eq.~(\ref{eq:ScH_est_simplified}).
It is worth noting that the parameters we used here only includes the material's properties,
without any fitting from direct MD simulations.

Figure~3 shows the comparison between the activation entropy (solid dots) from MEP calculations and the estimated values (dashed lines) from Equation~(\ref{eq:cross-slip:ScH_est}).
The materials properties from the interatomic potential~\cite{rao_atomistic_1999} are given as $K=\SI{183}{GPa}$, $\alpha_V=\SI{3.9e-5}{K^{-1}}$, and $\mu=\SI{81.4}{GPa}$.
The gradient of the shear modulus as a function of temperature is obtained from molecular dynamics of perfect crystal~\cite{clavier_computation_2017} as $\partial\mu/\partial T=\SI{1.77e-2}{GPa\cdot K^{-1}}$.
The perfect agreement indicates that the activation entropy can be well explained by the thermal expansion and thermal softening effects in the HTST with constant applied stress $\bm{\tau}_{\rm app}$.

\subsection*{VI. The determinant method for calculating the product of harmonic vibrational frequencies}

For the eigendecomposition of an \emph{non-singular} matrix $\mathbf{K}=\mathbf{V} \bm{\Lambda} \mathbf{V}^T$, the eigenmatrix is $\bm{\Lambda}={\rm diag}(\{\lambda_i,\,i=1,\dots,3N\})$, and the eigenvalues are $\nu_i\equiv\frac{\omega_i}{2\pi}=\frac{1}{2\pi}\sqrt{\frac{\lambda_i}{m}}$.
If we have the LU decomposition $\mathbf{K}=\mathbf{LU}$, the product of the eigenvalues can be written as,

\begin{align}
    \prod_{i=1}^{3N}\lambda_i &= \det(\mathbf{K}) = \det(\mathbf{L})\det(\mathbf{U}) \nonumber \\
        &= \prod_{i=1}^{3N}(L_iU_i)
    \label{eqn:cross-slip:product_by_LU}
\end{align}
where $L_i$ and $U_i$ are the diagonal elements of the $\mathbf{L}$ and $\mathbf{U}$ matrices.

After the Hessian matrix is modified with Eq.~(\ref{eq:modified_Hessian}),
the resulting matrix $\mathbf{K}$ becomes non-singular, and the product of the eigen frequencies can be obtained by the determinant (LU decomposition),

\begin{align}
    \prod_{j=1}^3\lambda^k_j\,\cdot\prod_{i=1}^{3N-3}\lambda_i = \det(\mathbf{K})
\end{align}
where $\nu^k_j$ are the three eigen values corresponding to the added spring forces.
It can be proved that the spring forces are independent to all the vibrational modes in the system.
The modified matrix $\mathbf{K}'$ can be written as,

\begin{align}
    \mathbf{K}' = \mathbf{K} + k\tilde{\mathbf{I}}_{3}
\end{align}
where $\tilde{\mathbf{I}}_{3}={\rm diag}(1,1,1,0,\dots,0)$ is a diagonal matrix with only three 1's on the main diagonal.
Since the diagonalization of $\mathbf{K}=\mathbf{V}\bm{\Lambda}\mathbf{V}^T$ satisfies $\mathbf{VV}^T=\mathbf{I}$,
we can write,

\begin{align}
    \tilde{\mathbf{I}}_3 = \mathbf{V}\tilde{\mathbf{I}}_3\mathbf{V}^T
\end{align}
Therefore, the modified matrix can be eigendecomposed into,

\begin{align}
    \mathbf{K}' = \mathbf{V}(\bm{\Lambda} + k\tilde{\mathbf{I}}_{3})\mathbf{V}^T
    \label{eqn:cross-slip:independent_k}
\end{align}
It is proved that the eigenvalues corresponding to the added spring forces $\lambda_j^k=k$.

The vibrational frequency term $\nu_{\rm HTST}$ is then calculated by the ratio between the determinants of states $A$ and $S$ since the frequencies from the spring forces are cancelled out:

\begin{align}
    \nu_{\rm HTST} &= \frac{\prod_{i=1}^{3N-3}\,\nu_i^{\rm A}}{\prod_{j=1}^{3N-4}\,\nu_j^{\rm S}} =
    \frac{\prod_{i=1}^{3N-3}\,\frac{1}{2\pi}\sqrt{\lambda_i^{\rm A}/m}}
         {\prod_{j=1}^{3N-4}\,\frac{1}{2\pi}\sqrt{\lambda_i^{\rm S}/m}} \nonumber \\
    &= \frac{1}{2\pi}\sqrt{\frac{\det(\tilde{\mathbf{K}}_A)\cdot\lambda^S_-}{\det(\tilde{\mathbf{K}}_S)\cdot m}}
\end{align}
where $\lambda^S_-$ is the negative eigenvalue that is obtained directly using the `eigs' method in MATLAB, and the determinant is calculated by LU decomposition in MATLAB.
In this work, we select $k = \SI{e-4}{eV/\angstrom}$ to avoid round-off errors.

\clearpage

\begin{figure}[!htb]
  \centering
  \includegraphics[width=\linewidth]{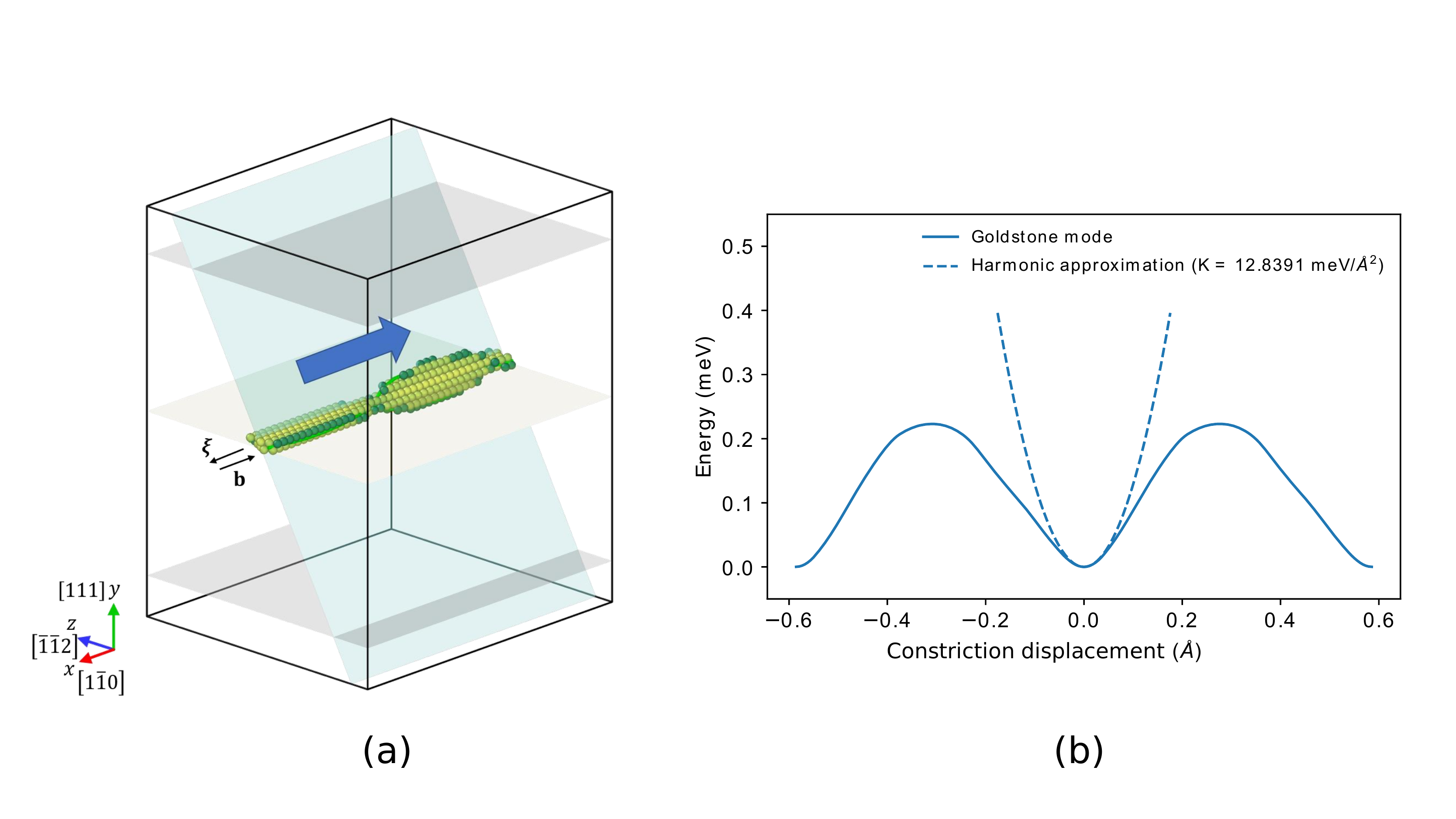}
  \label{fig:cross-slip:Goldstone_mode}
\end{figure}

\noindent {\bf Fig.~S1. Goldstone mode of the activated state}.
(a) Goldstone mode of the activated state S, the constrictions can move along the dislocation line without large frictions.
(b) Energy profile of the Goldstone mode and its harmonic approximation at the activated state. The horizontal axis is the total constriction displacement.

\clearpage

\begin{figure}[!htb]
  \centering
  \includegraphics[width=\linewidth]{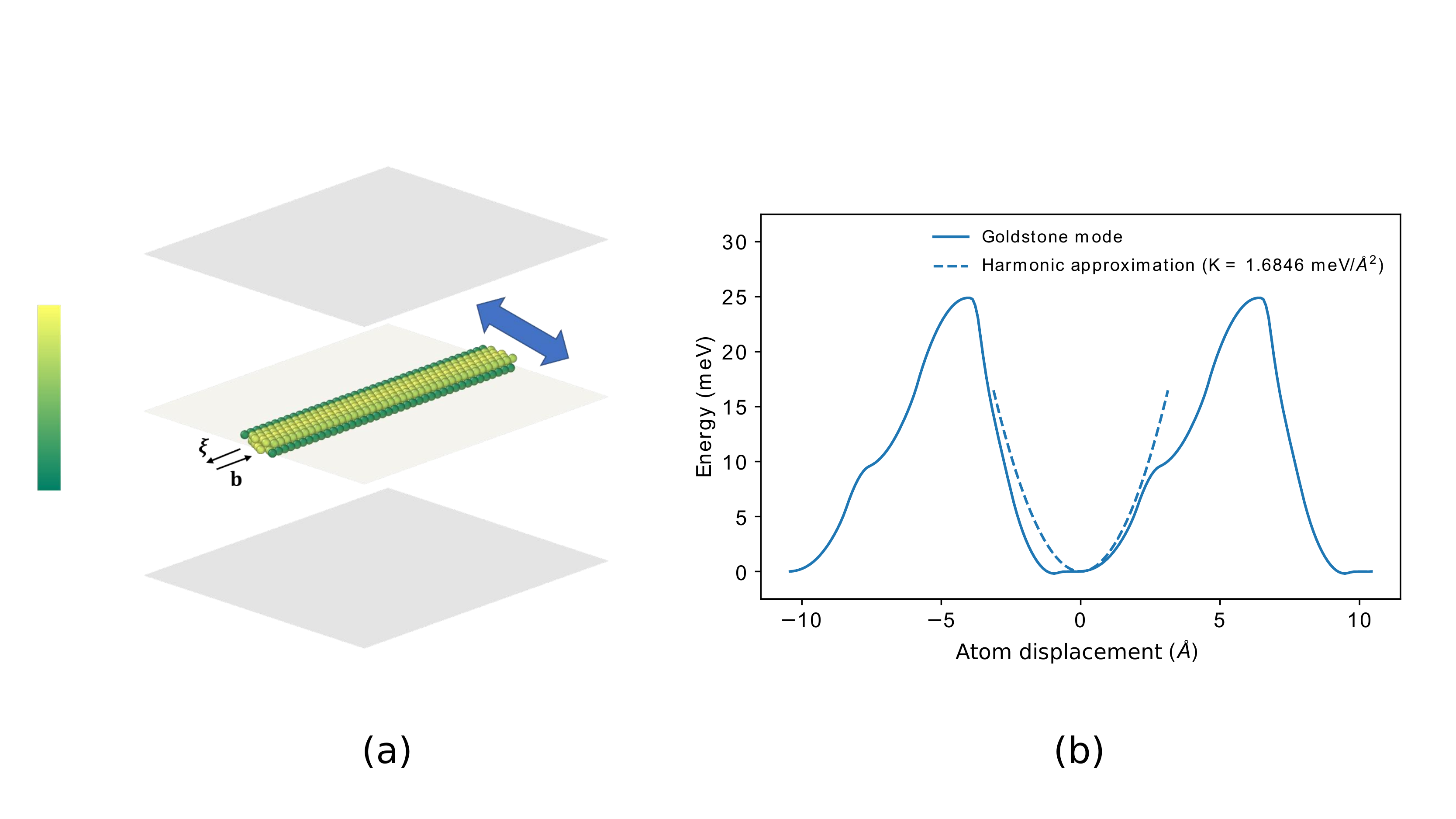}
  \label{fig:cross-slip:Gliding_mode}
\end{figure}

\noindent {\bf Fig.~S2. Gliding mode of the initial state}.
(a) Goldstone mode of the initial state $A$, the dislocation can glide along the original slip plane.
(b) Energy profile of the Gliding mode and its harmonic approximation. The horizontal axis is the total atom displacement.

\clearpage

\begin{figure}[!htb]
  \centering
  \includegraphics[width=\linewidth]{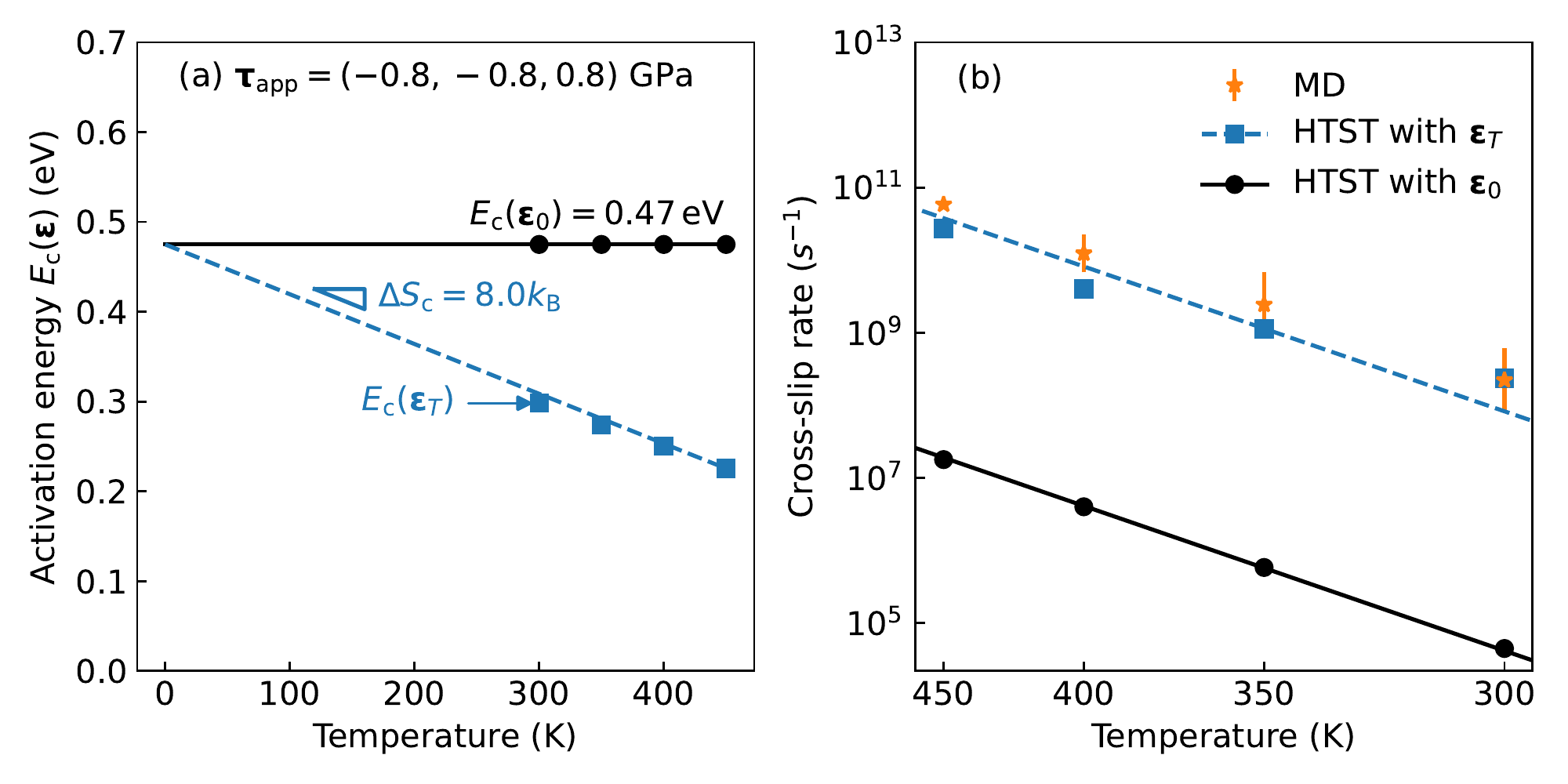}
  \includegraphics[width=\linewidth]{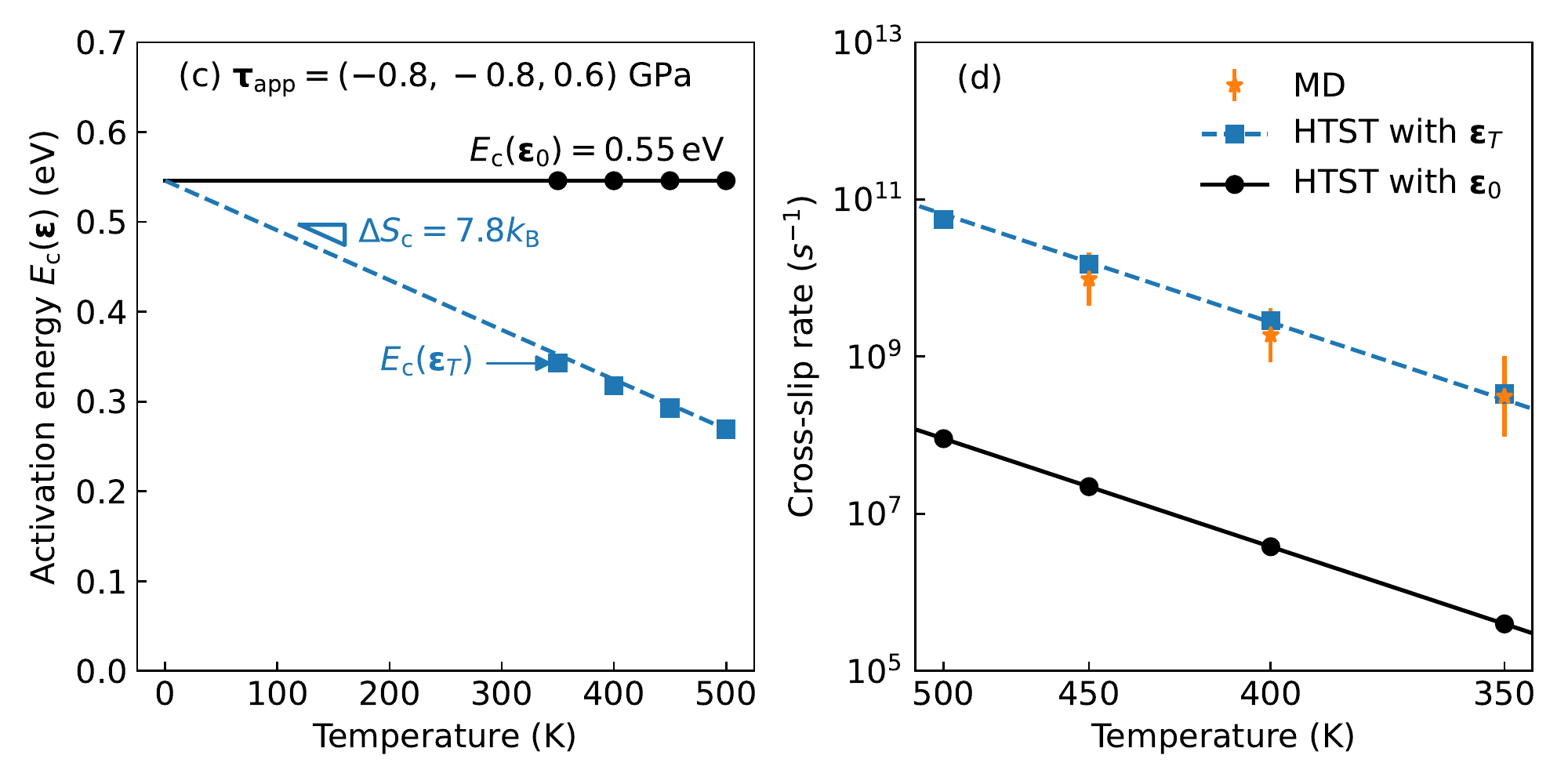}
  \label{fig:cross-slip:additional_cases}
\end{figure}

\noindent {\bf Fig.~S3. MD simulations and MEP calculations for additional applied stress conditions}.
(a) Activated energy calculated at zero-temperature strain $\bm{\varepsilon}_0$ and corresponding finite-temperature strain $\bm{\varepsilon}_{\rm T}$ for $\bm{\tau}_{\rm app}=(-0.8,-0.8,0.8)\si{GPa}$.
(b) Estimated rates using HTST (Eq.~(\ref{eq:rHTST})) with the activation energy and prefactor evaluated at $\bm{\varepsilon}_0$ and $\bm{\varepsilon}_{\rm T}$.
%
%
(c) Activated energy results and (d) Estimated rate results for the applied stress of $\bm{\tau}_{\rm app}=(-0.8,-0.8,0.6)\si{GPa}$.
The benchmark MD rates are performed for four temperature conditions ($T=350, 400, 450, \SI{500}{K}$) in (b) and three temperature conditions ($T=350, 400, \SI{450}{K}$) in (d), shown as the orange stars with error bar.

\clearpage

\begin{figure}[!htb]
  \centering
  \includegraphics[width=\linewidth]{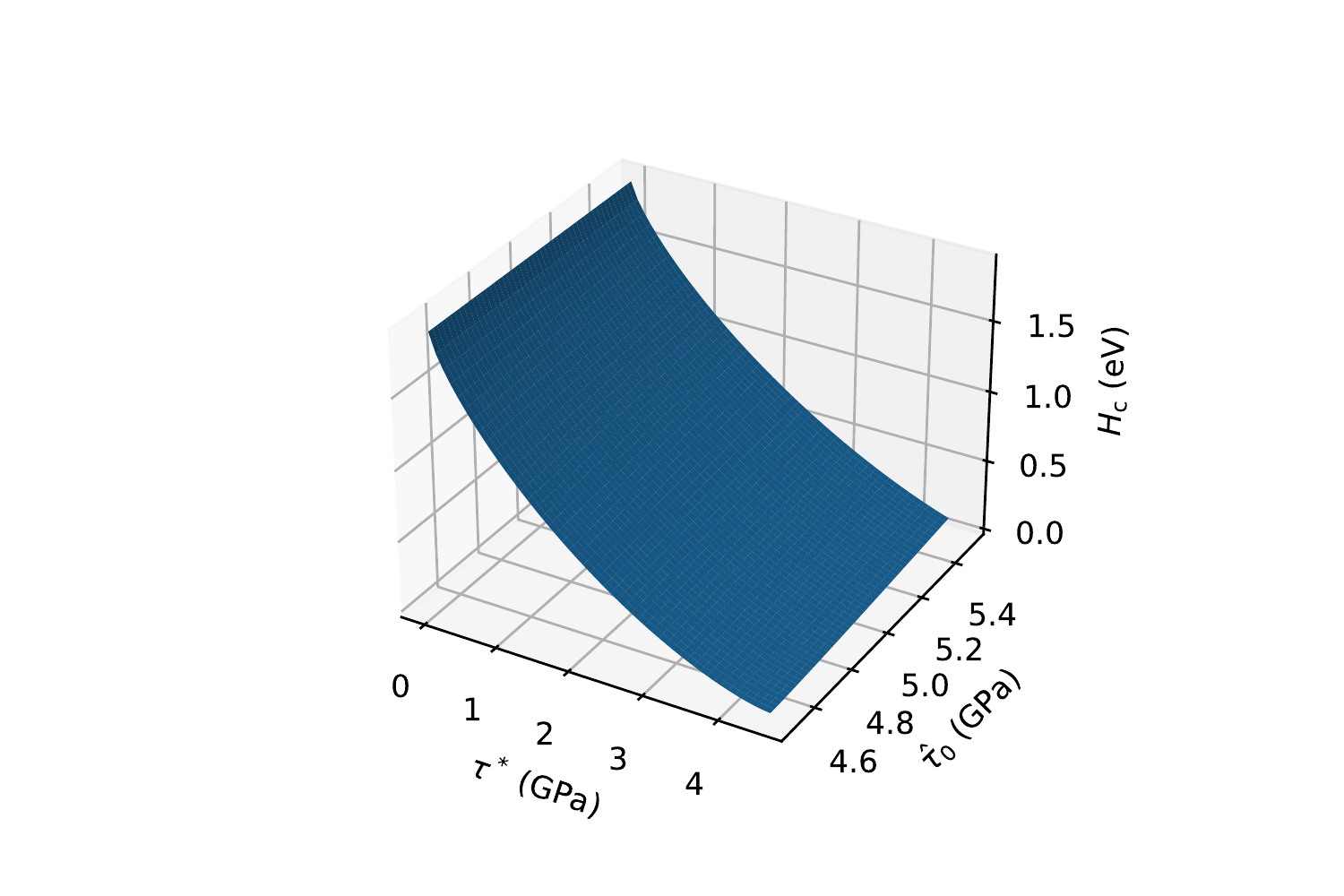}
  \label{fig:cross-slip:activation_enthalpy}
\end{figure}

\noindent {\bf Fig.~S4. Stress-dependent activation enthalpy}.
Activation enthalpy estimated from Eq.~(\ref{eq:cross-slip:activation_enthalpy}) as a two-dimensional function of $\tau^*$ and $\hat{\tau}_0$.

\clearpage

\begin{figure}[!hbt]
    \centering
    \includegraphics[width=0.9\linewidth]{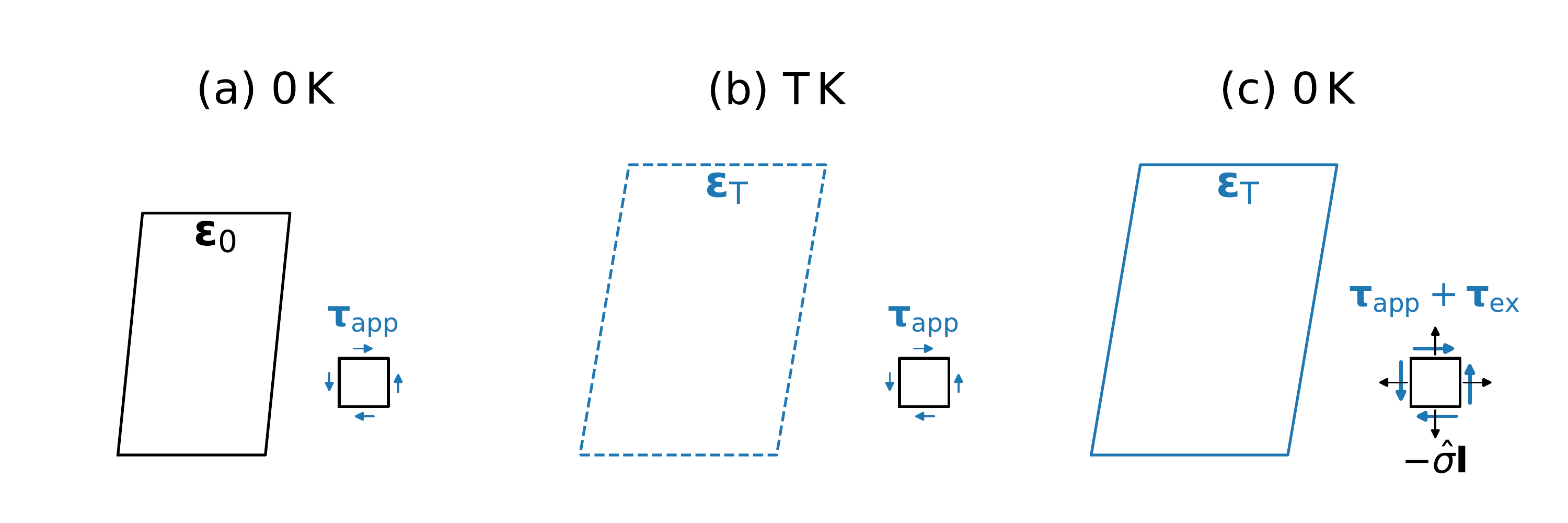}
\end{figure}

\noindent {\bf Fig.~S5. Thermal expansion and thermal softening effects}.
(a) Zero-temperature strain state $\bm{\varepsilon}_0$, with fixed applied stress $\bm{\tau}_{\rm app}$. 
(b) Finite-temperature strain state $\bm{\varepsilon}_{\rm T}$, with fixed applied stress $\bm{\tau}_{\rm app}$.
(c) Negative isotropic stress $-\hat{\sigma}\mathbf{I}$ and excess shear stress $\bm{\tau}_{\rm ex}$ are needed for reaching the finite-temperature strain state $\bm{\varepsilon}_{\rm T}$ at \SI{0}{K}.

\end{document}